\documentclass[aps,prl,twocolumn,showpacs,preprintnumbers]{revtex4-1}

\usepackage{psfrag,graphicx}
\usepackage{dcolumn}
\usepackage{amsmath,amssymb}
\usepackage{bm}
\usepackage{amsfonts,amssymb,amsmath}        % for math symbols.
\usepackage{epstopdf}
\usepackage{color}
\usepackage[dvipsnames]{xcolor}
\usepackage{microtype} %makes the text easier on the eye
\usepackage{braket} %typeset bras and kets

\newcommand{\be}{\begin{equation}}
\newcommand{\ee}{\end{equation}}
\begin{document}
\title{A proposal for a minimal surface code experiment}
\author{James R. Wootton, Andreas Peter, J\'{a}nos R. Winkler and Daniel Loss}
\affiliation{Department of Physics, University of Basel, Klingelbergstrasse 82, CH-4056 Basel, Switzerland}

\begin{abstract}

Current quantum technology is approaching the system sizes and fidelities required for quantum error correction. It is therefore important to determine exactly what is needed for proof-of-principle experiments, which will be the first major step towards fault-tolerant quantum computation. Here we propose a surface code based experiment that is the smallest, both in terms of code size and circuit depth, that would allow errors to be detected and corrected for both the $X$ and $Z$ basis of a qubit. This requires $17$ physical qubits initially prepared in a product state, on which $16$ two-qubit entangling gates are applied before a final measurement of all qubits. A platform agnostic error model is applied to give some idea of the noise levels required for success. It is found that a true demonstration of quantum error correction will require fidelities for the preparation and measurement of qubits and the entangling gates to be above $99\%$.

\end{abstract}


\maketitle

\section{Introduction}

Quantum error correction is the set of methods required to manage noise in a quantum computer \cite{lidar:13}. Before we can build a quantum computer we must first achieve quantum error correction, experimentally demonstrating that it can indeed correct quantum noise. In this paper we propose and study an experiment that could be used for this, and determine how low the noise levels must be in order for the experiment to succeed.

We consider the standard paradigm of quantum computation based on qubits --- two level quantum systems \cite{nielsen:00}. A \emph{physical qubit},
such as an electron-spin, interacts with its environment in intractable ways which we interpret as noise. Since noise must be kept arbitrarily low
for quantum computation, we encode one \emph{logical qubit}, i.e. the two level system we use for computation, in a highly entangled state of multiple physical qubits such that local noise can be traced
by measuring a set of observables  --- the stabilizers --- without affecting the logical qubit\cite{gottesman:96}.
A set of measurement results is called a syndrome. Interpreting a syndrome by means of a decoding algorithm shows possible error locations and thus allows
undoing errors on the logical state, but might also lead to solidification of the error's effect on the logical state if the physical errors 
were not properly identified. However, by increasing the number of physical qubits, the probability of failure can be made arbitrarily small.

\section{Experimental Development of Quantum Error Correction}

It can be expected that the experimental development of quantum error correction will consist of three distinct phases, after which full development of fault-tolerant quantum computation can begin.

Current experiments form the first phase, in which necessary primitives for quantum error correction are demonstrated. This includes showing that the required states can be prepared \cite{blatt:14}, that the required entangling measurements can be made \cite{martinis:14,chow:15,chow:16} and demonstrating correction of a subset of the errors \cite{martinis:14}. These experiments may also introduce artificial noise, to show the ability of the code to detect it \cite{blatt:14,chow:15}. Such experiments do not achieve quantum error correction themselves, but instead develop and demonstrate the necessary tools and techniques. This phase of development describes current experiments.

Future phases of proof-of-principle experiments will move on to larger codes (at least tens of qubits). They will use a complete set of syndrome measurements, capable of correcting all error types, and be operated at lower noise levels. This will allow them to fully achieve correction of the genuine noise experienced by the system.

We identify two future phases of such experiments. Experiments in the second phase will use only a single set of the required syndrome measurements before the final readout step. This is sufficient to detect and correct quantum errors in a way that will preserve the quantum information over a slightly longer time scale than without error correction.

In the third phase experiments the syndrome measurements will be repeated multiple times. This will allow quantum information to be preserved over much longer timescales. By using a number of measurement rounds that scales with system size, the lifetime should increase with system size also. This is the basis for arbitrarily increasing lifetime, as required for quantum computation.

Phase 2 experiments will be more straightforward than Phase 3 for several reasons. Firstly, the operations required for the syndrome measurement (entangling gates and ancilla measurements) need not be repeated. Secondly, the limited time frame in Phase 2 experiments allows less possibilities for logical errors to occur, which decreases their likelihood. The acceptable noise level will therefore be higher, and the number of qubits required to suppress the errors will be lower. Phase 2 should therefore be the current priority, in order to build up the techniques needed for Phase 3. In this work, we propose an experiment that could begin the second phase.

\section{17 qubit surface code}

\begin{figure}[t]
\begin{center}
{\includegraphics[width=8.5cm]{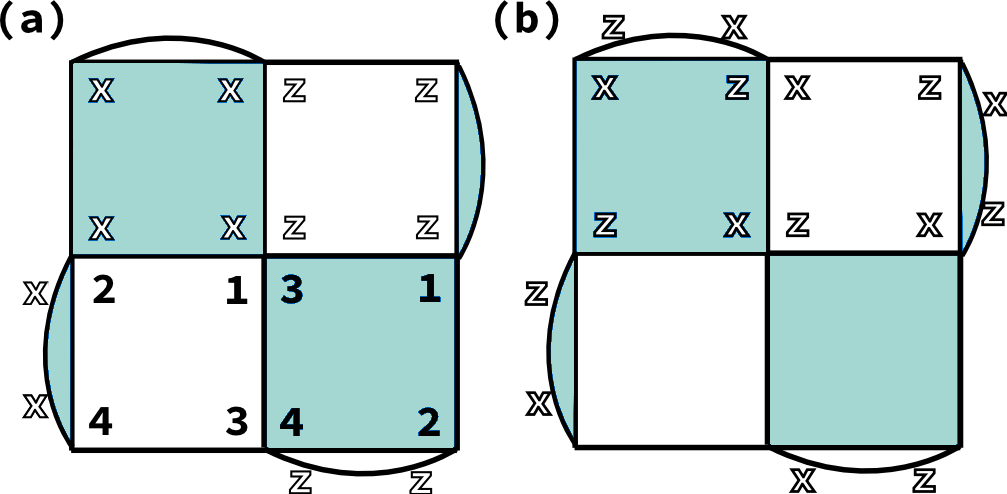}}
\caption{\label{code} Two alternative definitions of stabilizers in a surface code. Stabilizers are associated with plaquetes, inclduing the four semi-oval plaquettes on the edges. Code qubits reside on the vertices of the lattices, and ancilla qubits reside on the center of each plaquette.}
\end{center}
\end{figure}

\begin{figure}[t]
\begin{center}
{\includegraphics[width=6cm]{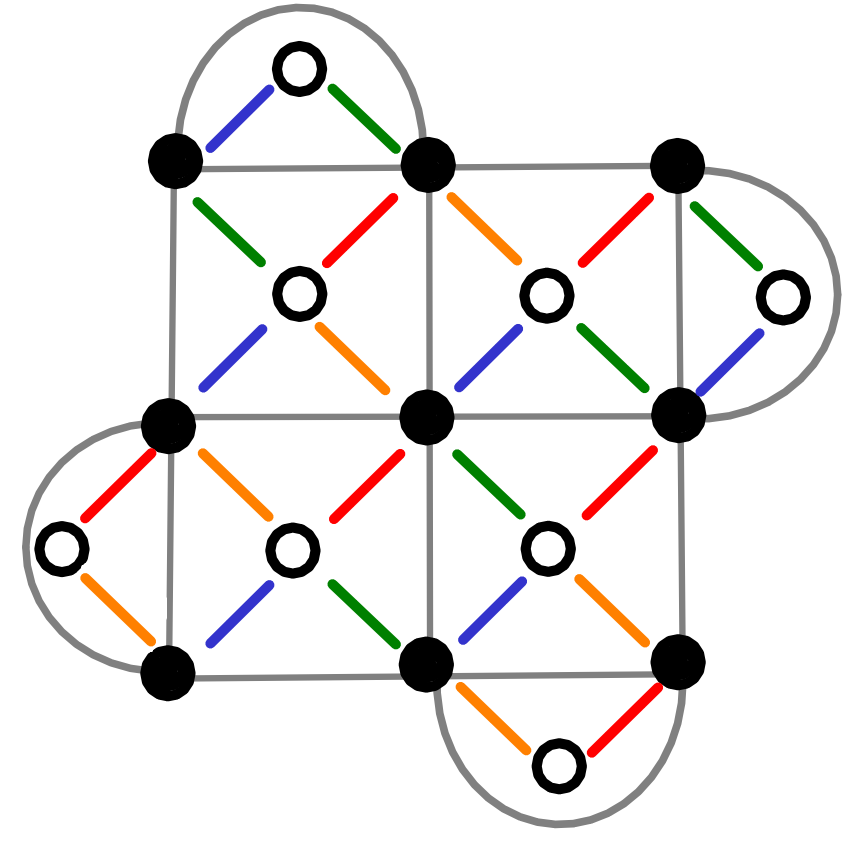}}
\caption{\label{colors} Code qubits are shown with filled circles, and ancilla qubits with empty circles. The entangling gates are shown with colored lines. The red gates are performed first, followed by orange, then green and finally blue.}
\end{center}
\end{figure}

The proposed experiment considers the simplest non-trivial instance of the surface code method of quantum error correction \cite{dennis}. It uses the smallest possible surface code that can both detect and correct quantum errors, which encodes a single logical qubit in 9 physical qubits \cite{svore:14}. These are known as the code qubits. Errors are detected through 8 stabilizer measurements. Each is measured using an additional physical qubit that acts as an ancilla. A total of 17 physical qubits are therefore required. The code is shown in Figs. \ref{code} and \ref{colors}.

The stabilizers of this code can be defined in multiple different ways. Two are shown in Figs. \ref{code}(a) and \ref{code}(b). The former allows a more straightforward explanation of the proposed experiment, and so will be used for the following description.

The stabilizer observables, used to detect noise, are associated with plaquettes of the lattice. These plaquettes are split into two groups: white and blue. The white plaquettes are made up of $\sigma_\textup{z}$ operators. They therefore detect the effects of $\sigma_\textup{x}$ and $\sigma_\textup{y}$ errors, which we call bit flips. The blue plaquettes are made up of $\sigma_\textup{x}$ operators and detect the phase flips $\sigma_\textup{z}$ and $\sigma_\textup{y}$. Note that
$\sigma_\textup{y}$ is composed of both a bit and a phase flip, and so is detected by both types of stabilizer.

Though each individual $\sigma_\textup{x}$ operation anticommutes with some of the white plaquettes, a string across the code from left to right commutes with all. We associate any such operation with the Pauli operator $X$ for the logical qubit. The eigenstates of these operators are therefore associated with the logical states $\ket{+}$ and $\ket{-}$. Similarly, a string of $\sigma_\textup{z}$ operations from top to bottom also commutes with all stabilizers and corresponds to the logical $Z$. Eigenstates of this are the logical qubit states $\ket{0}$ and $\ket{1}$.

Valid states of the logical qubit exist within the subspace that is the mutual $+1$ eigenspace of all stabilizers. Only highly entangled states exist within this space, and so their initialization will not be straightforward. However, consider the state for which all code qubits are $\ket{0}$. This is clearly in the $+1$ eigenspace of all white plaquette operators, and also for the logical $Z$ operation. It can therefore be associated with the logical $\ket{0}$ state. It is nevertheless not an eigenstate of the blue plaquette operators. Measurement of these will therefore yield completely random results.

After the measurement of all stabilizers, the state will be forced into a mutual eigenstate of all stabilizers. However, it will not be the mutual $+1$ eigenstate for the blue plaquettes, in general. To rectify this, $\sigma_\textup{z}$ operations could applied to a subset of the code qubits such that the state is rotated to the mutual $+1$ eigenstate. There are many ways in which this can be done, which will differ from each other by a logical $Z$, with no clear indication of which is preferred. The
procedure will therefore result in a logical $Z$ being applied randomly. However, since the initial logical state is $\ket{0}$, an eigenstate of $Z$, such an error would have no effect. We may therefore redefine our stabilizer space. Rather than the mutual $+1$ eigenstate of all plaquette operators, it will be the $+1$ eigenspace of white plaquettes, and whatever eigenspace is obtained by the first set of measurements for the blue plaquettes.

Similarly, the logical $\ket{1}$ state will be prepared when all code qubits are $\ket{1}$. The logical states $\ket{+}$ and $\ket{-}$ can also be prepared in a similar  way, but the roles of the white and blue plaquettes will be interchanged.

In general, initialization will be followed by a period in which stabilizers are measured periodically for an arbitrarily long time. Also, manipulations required for quantum computation will be applied between measurement rounds. The effects of noise over time will be detected by the stabilizer measurements. Combined with the use of an arbitrarily large surface code (defined on an arbitrarily large grid), this process will allow the lifetime of the logical qubit to be made arbitrarily long.

Each round of syndrome measurement is done using a five-step transversal process. Firstly, each ancilla is paired with a unique code qubit, and an entangling gate is applied between each pair. For the white plaquettes this will be a controlled-NOT, which either applies a $\sigma_\textup{x}$ or nothing to the ancilla depending on whether the code qubit state is $\ket{1}$ or $\ket{0}$, respectively. A similar gate is applied for the blue plaquettes, but controlled on the $\ket{+}$ and $\ket{-}$ states of the code qubits. This process is then repeated three more times, so that each ancilla is entangled with each of its neighbouring code qubits.

The pairing of the qubits is done according to the numbering in Fig. \ref{code}(a), and the coloring of Fig. \ref{colors}. All code qubits numbered $1$ are entangled first, and so on. The order is different for the two types of plaquette. This is to mitigate the effect of ancilla errors being spread to the code via the entangling gates, which can lead to uncorrectable errors if the ordering is not chosen carefully \cite{svore:14}.

The ancilla qubits are initialized as $\ket{0}$. The state of an ancilla after the entangling gates acts as a proxy for the multi-qubit plaquette-measurement, suffering
from additional noise due to imperfections in the entangling gates. The fifth and final step of the process is therefore to measure the ancilla qubits in the $Z$ basis. The result $\ket{0}$ implies that the code lies within the $+1$ eigenspace of the corresponding stabilizer, and $\ket{1}$ implies the $-1$ eigenspace.

These syndrome measurements must be constantly repeated until the time for the readout of the logical qubit. This is done in either the logical $Z$ or $X$ basis. The readout of logical $Z$ is done by measuring all code qubits in their $Z$ basis. This allows both the eigenstate of the logical $Z$ operator and a final syndrome measurement of the white plaquette stabilizers to be inferred. Logical $X$ readout is similarly done through measurement of the code qubits in the $X$ basis, and also yields a final syndrome for the white stabilizers. Once readout is complete, the combination of all syndrome measurements can be used to determined how to correct the value of the logical operator, in order to reflect the true state of the logical qubit.

The final syndrome measurement is different from standard ones in many important ways. Firstly, it can only be used to infer the measurement result for one type of plaquette: white plaquettes for $Z$ measurements and blue for $X$. However, since this is always the type of plaquette required for correction of the measured basis, it does not pose any problem.

Also the fact that the code qubits are measured directly means that noisy measurements have a different effect. Specifically, noise that causes the measurement to report the wrong value has an equivalent effect to a bit flip (for $Z$ measurements) applied directly before a perfect measurement. As such, we can consider this round to consist of perfect syndrome measurements, preceded by additional noise in the conjugate basis.

Since the ancilla qubits are not involved, there is another important difference. Standard syndrome measurements require use of the ancilla qubits at all points during the process: first for the entangling gates and then for measurement. It is therefore not possible to begin one round of measurements before previous one is finished. However, the code qubits are idle during the ancilla measurements. The readout measurements, which only involve code qubits, can therefore be performed concurrently with the ancilla measurements for the last standard syndrome round.

Finally, it is important to note that the readout measurements are not entangling. Applying readout directly after initialization would mean that the code never becomes entangled. At least one standard syndrome measurement is therefore required for the process to truly count as \textit{quantum} error correction.

By using the stabilizers of Fig. \ref{code}(a), we treat bit and phase flip errors in a completely equivalent but independent way. Such an approach would be fine if both occur at the same rate. However, typically there are large differences between their noise levels. Dealing with them independently therefore means that our error correction will always be constrained by the noisier of the two.

This can be dealt with using the stabilizers of Fig. \ref{code}(b) \cite{wen:plaquette}. These are equivalent to those of Fig. \ref{code}(a) up to local Hadamard rotations. All the analysis above therefore still applies, but with some exchanges between the $X$ and $Z$ bases of code qubits. This will lead to both types of stabilizers detecting both types of errors (though still only one type per code qubit). The noisier form of error will then be corrected more effectively, since they are always mixed with the less noisy ones. As such, it is the stabilizers of Fig. \ref{code}(b) that we consider in our proposed experiment.

\section{Proposed experiment}

For the simplest instance of surface code error correction, we need to implement the least number of syndrome measurement rounds. As noted above, applying readout directly after initialization does not lead to the state becoming entangled. However, it does yield a syndrome that can be used for error correction. This is effectively a classical error correcting code based upon the planar code. We therefore need to go beyond this by making a single standard syndrome measurement.

Just making this measurement is not enough. Consider the case for which the standard syndrome measurement is made, but the results are completely ignored. Error correction can be done using the readout syndrome alone. Due to the short timescale of the experiment, this may even provide good results. However, the nature of the error correcting problem would still be essentially classical. Applying the entangling gates for the standard syndrome measurement would have done little more than provide an additional source of noise. Even if they were essentially useless, the decoding could still succeed.

It is therefore important to ensure that the results of the standard syndrome measurement are indeed used, and that the error correction that results from their use is noticeably better than that when they are ignored. This will be an important condition that the results of an experiment must meet in order for it to be considered a success.

Given the use of a single standard syndrome measurement, the experiment will consist of the following steps. First the physical qubits are initialized in the required product state. This will correspond to an initial logical state of $\ket{0}$, $\ket{1}$, $\ket{+}$ or $\ket{-}$. The four rounds of transversal entangling gates will then be applied, as in Fig \ref{colors}. Finally, all physical qubits are measured. Ancillae are measured in the $Z$ basis, and each code qubit is measured in the $X$ or $Z$ basis as required. Note that, since one type of stabilizer has random outcomes due to initialization and is not measured during readout, its results will not be used in the error correction process. Only the two rounds of results for the other syndrome type are used. Specifically, for a $Z$ ($X$) measurement of the logical qubit we use only the outcomes of white (blue) stabilizers.

The basis chosen for readout of the logical qubit should always be the same as that of the initial state. With many samples, the fidelities of the states after error correction can then be determined.

For comparison, the same fidelity should also be determined for a single physical qubit. This allows us to assess the effectiveness of a code in protecting a logical qubit, in comparison with a logical qubit stored in a single physical qubit. It is only when this base level is improved upon that the error correction provides a benefit. This provides a second condition that our results must meet such that the experiment can be deemed to be a success.

\section{Decoding}

Decoding is the process determining how best to correct errors given the syndrome results. For the surface code, this is typically done using approximate \cite{fowler:rev,hutter:14,svore:14} or resource intensive algorithms \cite{wootton:high}. In either case, the nature of the noise model must be known for the most effective decoding. However, due to the small size of the proposed set-up, the use of such algorithms can be avoided.

The experiment consists of two rounds of syndrome measurement, of which only four stabilizers in each contribute to error correction. This means that there are only $2^8 = 256$ possible sets of measurement outcomes. Decoding can therefore be performed using a lookup table. Whenever we obtain a certain syndrome, we simply check the table to see whether keeping or flipping the readout logical value is more likely to give us the correct measurement result for the logical qubit.

The entries of this lookup table can be calculated using experimental data. The experiment can be run many times for different initial states. For each, the syndrome outcomes as well as the difference between the measured logical operator and the initial state is noted. The probability of the initial state being different to the measured logical operator can then be calculated for each syndrome value. With this data, we will know the best case strategy when we encounter each syndrome.

Note that no theoretical assumptions about the noise are required for our method. Also, the lookup table is calculated using the exact noise that affects the system. The correction applied will therefore depend on the exact details of the noise. It adapts to ensure that the most likely correction is always applied.

For comparison, we also consider the decoding method proposed by Tomita and Svore in \cite{svore:14}. Like other algorithmic methods for surface codes, this uses the fact that a sequence of nearby errors can form a so-called \emph{error-chain}. These result in syndrome changes only at any endpoints of the chain that reside in the bulk. Decoding can therefore be done by pairing syndrome changes with each other or an edge.

The Tomita-Svore method also uses the fact that any syndrome change can be caused by no more than a single error. This allows the minimal number of errors that causes a given syndrome to be determined using a simple set of rules. The results are equivalent to the minimum weight perfect matching algorithm applied with all errors assigned the same weight. For convenience and due to the limited code-size, the algorithm can be cast into a lookup-table. This would link each syndrome to a set of likely errors, and their corresponding effect on the readout of the logical operator. We have adapated the method slightly to account for the fact that only two measurement rounds are performed in the proposed experiment. Specifically, any syndrome defects that remain after the Tomita-Svore rules are applied are assumed to be due to errors at the edge.

\section{Determining Threshold Noise Levels}

An important consideration for proof-of-principle experiments is how low the noise level must be. Our main focus is on determining noise levels that will allow the surface code to outperform a single physical qubit.

Determining thresholds usually requires the application of a decoding algorithm to a large and representative set of noise samples. However, the use of a full look-up table decoder allows us to avoid this. Instead we calculate two probabilities for each syndrome: the probability that the value of the read out logical operator has been affected by the noise, and the probability that it has not. Decoding will be done according to the most likely of these two options, given the measured syndrome. The probability of successful decoding for a given syndrome therefore corresponds to the larger of the two. To obtain the total probability of successful decoding one can simply sum these, weighted by the probability of the corresponding syndromes. We refer to this total success probability as the fidelity of the process.

\subsection{Simulating Noise}

Quantum noise is, in general, complex and difficult to simulate. In studies of error correction it is common to approximate noise by using Pauli channels. These are described by the application of the Pauli operators $\sigma_\textup{x}$, $\sigma_\textup{y}$ and $\sigma_\textup{z}$ to qubits within the system with certain probabilities. The application of these operations on a stabilizer code, such as the planar code, always results in a definite change in the values of the syndromes and logical operators. It is therefore simple to simulate, and can be done using standard methods for Clifford circuits \cite{nielsen:00}.

Unfortunately, this simplicity comes at a cost. The approximation to Pauli channels means that coherent effects of noise are not taken into account, which can lead to significantly different results in some cases \cite{darmawan:16}.

Another approach is to use a full simulation of noise that cannot be expressed as a Pauli channel, but must instead be treated as a more general noise channel. Due to the relatively small nature of our system, this can be done using a \emph{tensor network} simulation. 
In the tensor network approach, the density operator of the system is
decomposed into tensors that each describe a physical qubit and its
entanglement with neighboring qubits. 
Gates can then be applied exactly to one or two qubits by operating on their
tensors, without having to take the whole density operator into account 
at each step.

Following the approach of \cite{darmawan:16}, we factorize  the initial state into a \emph{Projected Pair Entangled State} (PEPS) and then apply single- or two qubit channels to reflect the operations of local noise and the (noisy) application of entangling gates respectively. Due to the limited number of entangling gates and the inherently limited complexity of a two-qubit channel, the exact reproduction of the algorithm is possible. Given this exact representation
of the density operator, we can calculate the expectation values of all
syndromes to create a lookup-table that is accurate up to numerical
inaccuracies. 

In this work we will consider both general noise channels and their Pauli approximations. Tensor network simulations will be used to calculate results in both cases. The results for Pauli channels will allow us to comment on the connection with many existing results in the error correction literature that use this method. The results for the general noise channels will be more realistic and incorporate the coherent effects of the noise.

\subsection{General features of the noise model}

To determine the noise model to be considered, let us first consider the noise experienced by a single physical qubit that is not involved in a surface code. The fidelity with which this can store a logical qubit is the benchmark that the code must beat.

When preparing the physical qubit in the required initial state, $\ket{0}$, $\ket{1}$, $\ket{+}$ or $\ket{-}$, there will be some probability that this preparation fails. We model this as a perfect preparation followed by a bit flip with probability $p_\textup{bit}$ and a phase flip with probability $p_\textup{phase}$. The former affects preparation of the $Z$ basis states, swapping one for the other, and the latter affects the $X$. For simplicity we will assume that $p_\textup{bit} =
p_\textup{phase} = p$.

The qubit will then experience decoherence for a time $t$. For later convenience, we consider this as four successive periods of time $t/4$, the time for each entangling gate implemented in the surface code experiment. We consider the decoherence to be characterized by two timescales, $T_1$ and $T_2$. The former is the timescale for amplitude damping, and the latter for dephasing.

Finally, the qubit is measured in the $Z$ or $X$ basis. This measurement will be imperfect, giving the opposite result with some probability. We model this as a perfect measurement preceded by a bit flip with probability $m_\textup{bit}$ and a phase flip with probability $m_\textup{phase}$. The former leads to the incorrect outcome for a $Z$ measurement, and the latter for an $X$ measurement. For simplicity we will assume that $m_\textup{bit} = m_\textup{phase} = m$.

We are interested in the probability that the qubit is measured in the same state in which it was prepared, which we call the fidelity of the single qubit storage process. There will be two separate fidelities $F_\textup{single}^\textup{z}$ and $F_\textup{single}^\textup{x}$, for the two bases used. The overall fidelity is taken to be the minimum of these: $F_\textup{single} = \min(F_\textup{single}^\textup{z},F_\textup{single}^\textup{x})$.

Now we consider the noise applied to physical qubits used within the surface code. This begins with preparation noise and ends with measurement noise, as described above. 

During the four rounds in which entangling gates are applied, the noise of these gates must be considered. The nature of such noisy gates will depend strongly on the physical system used to implement them. Also, these gates will typically be constructed from a sequence of multiple single- and two-qubit rotations. The exact choice of this sequence will also determine the form which the noise will take.

It is beyond the scope of this work to consider such platform specific details. Instead we will look at the kind of error model used currently for benchmarking quantum error correction. This will then give an idea of what is required for specific platforms, while also being connected to the many previous threshold results in the field. Specifically, we model imperfect two-qubit gates by applying depolarizing noise independently to the two qubits involved. The manner in which this is done depends strongly on the kind of error channel used, and so will be described further below.

Due to the nature of the stabilizers at the edge, not all qubits are involved in a two-qubit gate at all times. Such idle qubits will experience decoherence for the corresponding time $t/4$.

Note that the interactions causing decoherence in the idle qubits will also be applied to those involved in the two-qubit gates. Indeed, this is one of the effects that will contribute to their imperfection. As such, we must ensure that no tests are made such that the qubits involved in these gates experience less noise than the idle ones.

The total fidelity for this process is taken to be the probability that the logical state of the initial configuration can be recovered
given the syndrome. This will yield two fidelities, $F^\textup{z}$ and $F^\textup{x}$, for the two bases used. The overall fidelity is taken to be the minimum of these: $F = \min(F^\textup{z},F^\textup{x})$. The fidelity for the code when the decoding uses both syndrome measurement rounds will be denoted $F_\textup{code,2}$. That for only the final round will be $F_\textup{code,1}$.

\subsection{Pauli noise channels}

Pauli noise channels apply only errors of the form $\sigma_\textup{x}$, $\sigma_\textup{y}$ and $\sigma_\textup{z}$. The channel is described by the probabilities for each of these processes.

The effects of the amplitude damping and dephasing noise can not be fully captured by Pauli channels, but it can be approximated as one. This applies $\sigma_\textup{x}$, $\sigma_\textup{y}$ and $\sigma_\textup{z}$ errors with the following probabilities \cite{svore:14}
\be
d_\textup{x} = d_\textup{y} = \frac{ 1-e^{t/(4T_1)} }{4}, \,\,\, d_\textup{z} = \frac{ 1-e^{t/(4T_2)} }{2} - d_\textup{x}.
\ee
The total probability of an error is therefore $d=d_\textup{x}+d_\textup{y}+d_\textup{z}$. 

Imperfect entangling gates are modelled with a Pauli channel by applying depolarizing noise immediately prior to a perfect two qubit gate. The noise is independently applied to each qubit, with each Pauli error occurring with probability $g/3$. Here $g$ denotes the total probability of depolarizing noise for each qubit. For these Pauli channels, the condition that qubits involved in entangling gates should not experience less noise than idle ones corresponds to ensuring that $d \leq g$.

\subsection{General noise channels}

In the tensor network, all operations are represented by tensors 
$\mathcal{E}_{ijj'i'}$
that are derived from quantum channels $\mathcal{E}$ as 
\be
\mathcal{E}_{ijj'i'} := \bra{i} \mathcal{E} \left( \ket{j}\bra{j'} \right) \ket{i'},
\ee
where $\ket{j},\ket{j'}$ and $\ket{i}, \ket{i'}$ are the bases in terms of which the input and output are expanded.
The channels themselves are calculated by numerically solving the 
Lindblad-equation. For the amplitude and phase damping noise, the Lindblad-operators are $\sqrt{\gamma} \sigma_+$, $\sqrt{\gamma} \sigma_-$ and $\sqrt{\phi} \sigma_\textup{z}$. For the depolarizing noise applied during the entangling gates, the Lindblad-operators are $\sqrt{\omega} \sigma_i$ with $i\in {x,y,z}$. The values of $\gamma$, $\phi$ and $\omega$ are
\begin{align} \nonumber
\gamma &= \tfrac{1}{16 T_1}, \\ \nonumber
\phi &= \tfrac{1}{2} \left( \tfrac{1}{8 T_2} -\gamma \right), \\
\omega &= -\tfrac{1}{8} \log \left[1- \tfrac{4}{3} g\right].
\end{align}
These values ensure that the solution of the Lindblad-equation for a trivial Hamiltonian corresponds to the noise described above.

\subsection{Comparison to other thresholds}

The standard means to calculate thresholds for surface codes is to look at asymptotic properties. The threshold is defined as the noise level below which the logical error rate decays with increasing system size. See \cite{fowler:rev,wootton:high,anwar:14} for examples. This threshold is equivalent to a phase transistion in corresponding statistical mechanics models \cite{dennis,andrist:15}. Furthermore, the requirements of large scale simulations mean that the noise considered is typically in the form of Pauli channels.

These results give a value for the $p=m=g$ threshold of around $1\%$ when arbitrarily many measurement rounds are used \cite{fowler:rev}. For the case of a single round with perfect measurements ($g=0$ and $p=m$ nonzero only for code qubits), the threshold is $p+m=11\%$ \cite{dennis}.

In our case we investigate the required noise levels for a small scale experiment. The definition of the threshold is therefore quite different. We cannot expect to directly apply results from studies with the above methodology. It will therefore be interesting to see to what extent they agree.

Note that our use of only two measurement rounds means our case is more similar to that with a threshold of $p+m=11\%$ than that with $p=m=g=1\%$. We can therefore expect that the threshold will decrease as more measurement rounds are added to an experimental setup. The noise levels suitable for our proposal must therefore still be improved upon before more complex experiments can be attempted.

\section{Results}

\subsection{Preliminaries}

For the time scales, $T_1$, $T_2$ and $t$, we note that the absolute values do not matter, only their relative durations. All times are therefore stated in units of $T_2$. The remaining free parameters are the timescales $T_1/T_2$ and $t/T_2$, and the probabilities $m$, $p$ and $g$.

With given values of these parameters, we can now caculate the fidelity for the error correction process and determine whether or not the experiment would be successful in demonstrating quantum error correction. Specifically for fixed values the timescales $T_1/T_2$ and $t/T_2$ we find the highest $g$ that would lead to success for various values of $m$ and $p$. This is then repeated for different values of the ratios.

Here success means that the code should outperform a single qubit ($F_\textup{code,2}>F_\textup{single}$) and that the error correction when both syndrome rounds are used is significantly better than that for only the last ($F_\textup{code,2} \gg F_\textup{code,1}$). To determine the degree to which the latter is satisfied, we consider the quantity
\be \nonumber
f = \frac{ 1-F_\textup{code,2} }{ 1-F_\textup{code,1} }.
\ee
A value of $f<1$ here demonstrates an advantage in using both syndrome measurement rounds. If the measured value of $f$ cannot be distinguished from unity within the precision of the data, no such advantage is demonstrated.

In the results we will see that reasonable noise levels achieve values of $f$ no lower than around $90\%$. Decoding using both rounds is therefore demonstrably better than just using the final one. However, the difference is not huge.

We can use this fact to further simplify preparation and measurement noise. Consider the case where only the results from the final round of syndrome measurement is used. The difference between errors made before and after the entangling gates is effectively removed in this case, because the issue regarding whether they'll be seen by the standard syndrome measurement is not relevant. The preparation and measurement noise can therefore be effectively considered as a single noise type with a combined strength of $p(1-m)+m(1-p) \approx p+m$. The results will then not depend strongly on whether this noise is distributed evenly between the two processes, or biased all on one. As such we can restrict to $p=m$ in this case without loss of generality.

For cases in which results from both rounds of syndrome measurements are used, this equivalence between $p$ and $m$ no longer holds. However, as long as the addition of the results from the standard syndrome measurement round does not have a very strong effect, the equivalence will still hold approximately. Given the values of $f$ that we consider, we can therefore continue to use $p=m$.

\subsection{Numerical results}

It can be expected that the threshold value of $g$ will have a complex relationship with $p=m$. On the one hand, lowering $p=m$ will allow more effective error correction, and so could allow good results even for larger $g$. On the other hand, lowering $p=m$ also reduces the amount of noise felt by the single qubit with which we compare. The requirement for success therefore becomes more stringent, which may lead to a lower $g$ being required.

\begin{figure}
\includegraphics[width=\columnwidth]{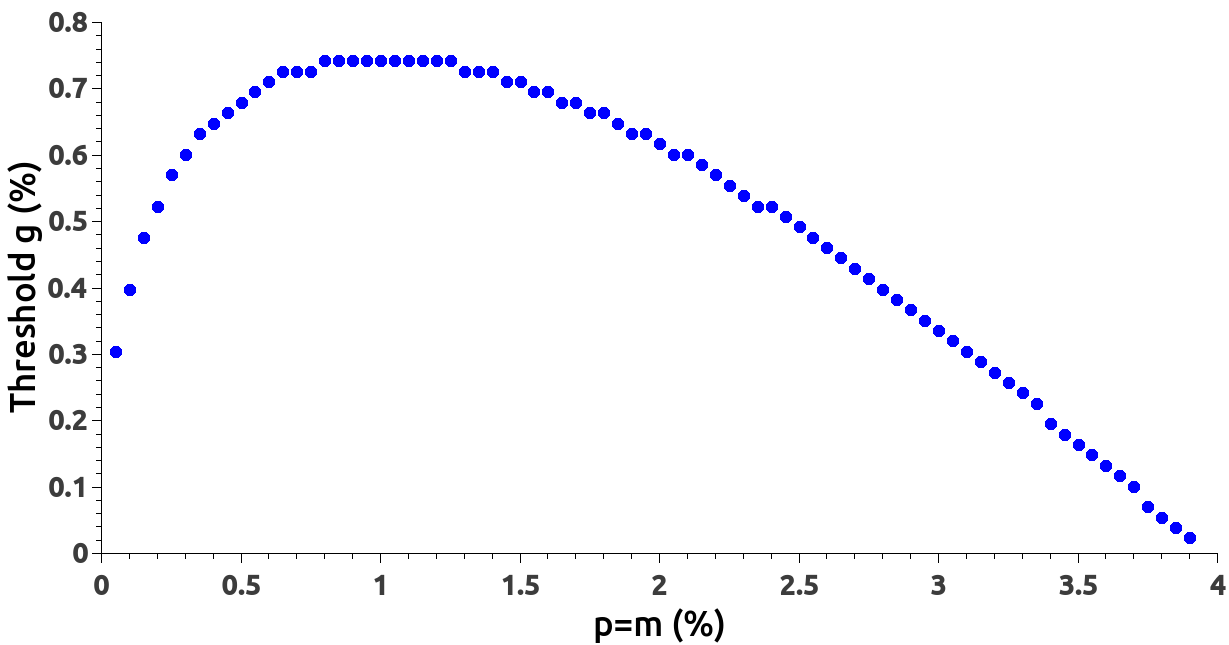}
\caption{\label{threshold} Graph of the threshold for gate noise, parametrized by the probability $g$, against preparation and measurement noise, parametrized by $p=m$.}
\end{figure}

\begin{figure}
\includegraphics[height=6cm]{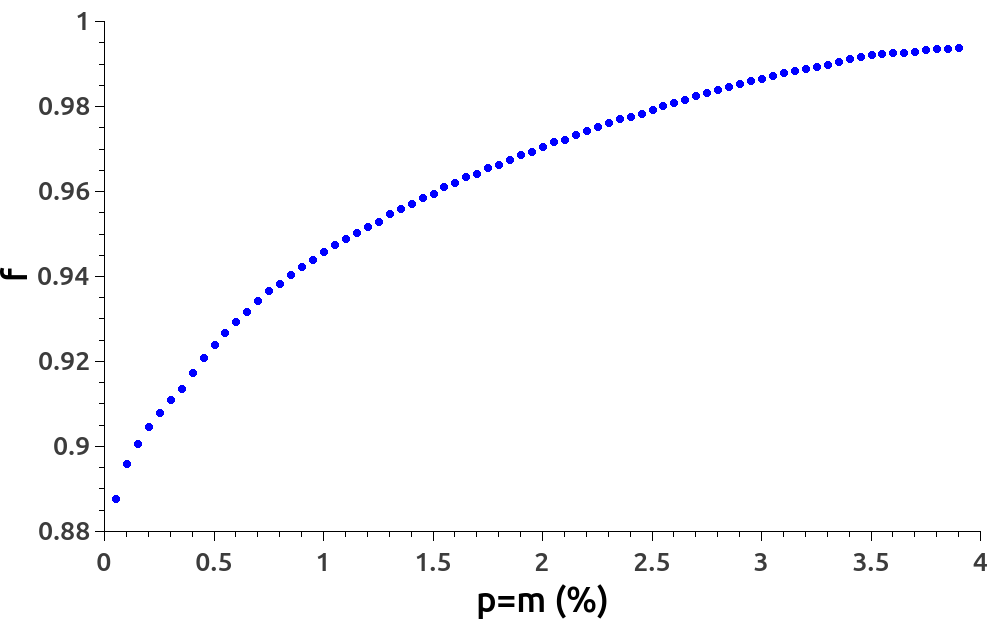}
\caption{\label{ratio} Graph of the ratio $f$ achieved at the threshold value $g$ for different preparation and measurement noise.}
\end{figure}

Our numerical results demonstrate the trade-off between these two effects. The the threshold value of $g$ is plotted against $p=m$ in Fig. \ref{threshold} for the case of $T_1/T_2 =10^4$ and $t/T_2 = 10^{-3}$ for the general noise channel. The corresponding values of $f$ are shown in Fig. \ref{ratio}. The graphs for $T_1/T_2 =10^2$ and $T_1/T_2 =10^3$ are essentially identical, showing that the results do not depend strongly on these timescales when the entangling gates are much faster than the decoherence times. The graphs for the noise approximated to a Pauli channel are also essentially identical. The forms of noise considered are therefore ones for which the Pauli approximation is well justified.

The full results for the threshold show that the two effects described above lead to a maximum threshold for the gate noise of around $g=0.7\%$. This occurs at around $p=m=1.25\%$ and achieves a respectable value of $f=0.95\%$ for the ratio. This is possibly the most experimentally amenable set of noise levels that allow for the demonstration of quantum error correction in the proposed experiment.

If preparation and measurement noise are lowered, the threshold for $g$ is lowered accordingly. At the extreme case of making preparation and measurement perfect, the threshold for the entangling gates is $g=0.03\%$ and $f<0.9$. Moving in this direction would therefore allow an even more effective demonstration of quantum error correction.

If preparation and measurement noise are raised, success in the experiment is found to be impossible beyond $p=m=4\%$. At this point, the entangling gates must be essentially perfect. However $f=1$ in this case, and so the situation is effectively one of classical error correction. Though the entangling gates are perfect, their results do not help provide better error correction. Such a noise regime is therefore not useful for the proposed experiment.

\begin{figure}
\includegraphics[height=6cm]{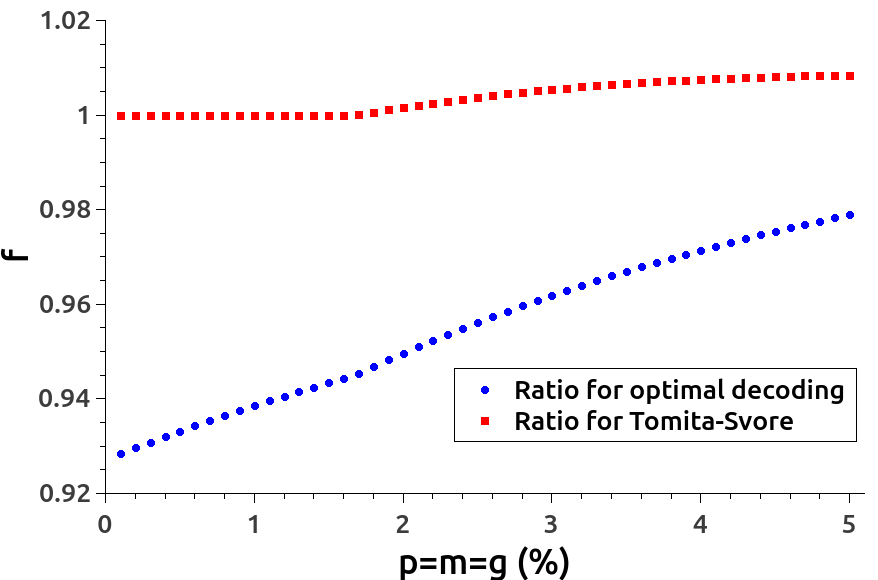}
\caption{\label{TS_comparison} Graph of the ratio $f$ for different values of the noise rate $p=m=g$. Results are shown for both the optimal decoder (a lookup table populated by experimental data) and Tomita-Svore (a lookup table based on a set of rules).}
\end{figure}

The threshold results for the Tomita-Svore decoder are largely similar to that for the optimal decoding. The value of the ratio $f$, however, shows much larger differences, as seen in Fig. \ref{TS_comparison}. In this figure we look at data for the case of $p=m=g$. The ratio $f$ is plotted for both the optimal decoding and Tomita-Svore decoding. In both cases the denominator of the ratio is the $F_\textup{code,1}$ obtained from optimal decoding. This is to ensure that both compare against the
same thing, and because the simplicity of decoding for the single round syndrome means that optimal decoding can always be used for $F_\textup{code,1}$.

It is found that the advantage of using the full syndrome is dramatically reduced when Tomita-Svore decoding is used. In fact, results are worse than for single round decoding. This is quite surprising, since this decoder gives essentially identical curve as Fig. \ref{threshold} for the threshold values of $g$. Demonstrating the advantage of using the full syndrome therefore has a strong need for optimal decoding. Even near optimal decoding can make success much harder to achieve.

\section{Conclusions}

We have proposed a minimal surface code experiment, which would be able to form the first demonstration of quantum error correction. For the noise model considered, it was found that the fidelities for preparation and measurement noise should be better than around $99\%$. Imperfections in the entangling gates were parametrized by the probability $g$ for single qubit depolarizing noise. It was found that this should be less than around $0.7\%$. This corresponds to a fidelity of around $99\%$ for the entangling gates as a whole. These results were found for both a Pauli approximation of noise, and a full simulation including coherent effects.

The results agree well with previous results using Pauli noise and with thresholds defined as phase transitions for codes of arbitrarily large size and arbitrarily many measurement rounds. However, it is significantly less than that for only a single measurement round. This suggests that the acceptable noise rate for more complex experiments in future will require fidelities beyond the $99\%$ level.

It is found that success for this experiment strongly depends on the use of an optimal decoder. For more advanced experiments, with many syndrome measurement rounds, it can be expected that the Tomita-Svore decoder will provide good results. However, it is not clear what decoding should be used for experiments for a medium number of measurement rounds. These will be beyond the complexity for which our method can be efficiently applied, but may not yet be in the regime for which Tomita-Svore decoding excels. Methods should therefore be designed to address this need, achieving near optimal decoding tailored to the noise model within a reasonable time scale. Possibilities could be an adaption of the Markov chain Monte Carlo decoding of \cite{wootton:high,hutter:14}, or the use of a genetic algorithm \cite{weasel}.

\section{Acknowledgements}

The authors would like to thank Christoph Kloeffel for discussions and the SNSF and NCCR QSIT for support.

\bibliography{refs}

\end{document}